\documentclass[12pt]{article}
\usepackage{latexsym, amssymb, graphics, amsmath}
\newcommand{\simge}{\hspace*{0.2em}\raisebox{0.5ex}{$>$}
     \hspace{-0.8em}\raisebox{-0.3em}{$\sim$}\hspace*{0.2em}}

\newcommand{\beq}{\begin{equation}}
\newcommand{\eeq}{\end{equation}}
\newcommand{\beqa}{\begin{eqnarray}}
\newcommand{\eeqa}{\end{eqnarray}}
\newcommand{\boldtau}{\mbox{\boldmath $\tau$}}

\newcommand{\boldpi}{\mbox{\boldmath $\pi$}}
\newcommand{\slashT}{\slash \hspace{-0.4em}T}

\begin{document}

\begin{titlepage}

\vspace{2.5cm}

\begin{center}
{\Large\bf The Electric Dipole Form Factor of the Nucleon\\
in Chiral Perturbation Theory to Sub-leading Order}

\vspace{2.0cm}

\renewcommand{\thefootnote}{\ensuremath{\fnsymbol{footnote}}}

{\bf E. Mereghetti}$^1$
,
{\bf J. de Vries}$^2$
,
{\bf W.H. Hockings}$^3$
,\\
{\bf C.M. Maekawa}$^4$
, and
{\bf U. van Kolck}$^1$

\vspace{0.8cm}
$^1${\it Department of Physics, University of Arizona}

{\it Tucson, AZ 85721, USA}

$^2${\it KVI, Theory Group, University of Groningen}

{\it 9747 AA Groningen, The Netherlands}

$^3${\it Department of Mathematics and Natural Sciences, 
Blue Mountain College}

{\it Blue Mountain, MS 38610, USA}

$^4${\it Instituto de Matem\'atica, Estat\'{i}stica e F\'{i}sica, 
Universidade Federal do Rio Grande}

{\it Campus Carreiros, PO Box 474, 96201-900 Rio Grande, RS, Brazil}

\end{center}

\vspace{1.5cm}

\begin{abstract}
The electric dipole form factor (EDFF) of the nucleon stemming from
the QCD $\bar{\theta}$ term and from
the quark color-electric dipole moments is calculated in
chiral perturbation theory to sub-leading order.
This is the lowest order in which the isoscalar EDFF
receives a calculable, non-analytic contribution from the pion cloud.  
In the case of the $\bar{\theta}$ term,
the expected lower bound on the deuteron electric dipole moment is
$|d_d|\simge 1.4 \cdot 10^{-4} \bar\theta \, e$ fm.
The momentum dependence of the isovector EDFF is proportional to a
non-derivative time-reversal-violating pion-nucleon coupling,
and the scale for momentum variation
---appearing, in particular, in the radius of the form factor---
is the pion mass.
\end{abstract}

\vspace{2cm}
\vfill
\end{titlepage}

\setcounter{page}{1}

The electric dipole form factor (EDFF) completely
specifies the parity ($P$) and time-reversal ($T$) -violating coupling 
of a spin 1/2 particle to a single photon \cite{WHvK,willthesis}.
At zero momentum, it reduces to the electric dipole moment (EDM), 
and  its radius provides a contribution to the Schiff moment (SM)
of a bound state containing the particle \cite{scott}.
The full momentum dependence of the form factor 
can be used in lattice simulations to extract the EDM by extrapolation 
from a finite-momentum calculation \cite{wilcox}
(in addition to the required extrapolations in quark masses
and volume \cite{Savage}).

There has been some recent interest 
\cite{WHvK,willthesis,Faessler,otherFaessler,Ottnad,jordyetal}
in the nucleon EDFF motivated by prospects of experiments that aim
to improve the current bound on the neutron
EDM, $|d_n| < 2.9 \cdot 10^{-13} \, e$ fm \cite{dnbound},
by nearly two orders of magnitude \cite{expts},
and to constrain the proton and deuteron
EDMs at similar levels \cite{storageringexpts}.
We would like to understand the implications of
a possible signal in these measurements to the sources
of $T$ violation at the quark level,
which include, in order of increasing dimension, the QCD $\bar\theta$ term, 
the quark color-EDM (qCEDM) and EDM, the gluon color-EDM, 
{\it etc.} \cite{Rujula,Weinberg:1989dx}.
Unfortunately, 
as with other low-energy observables, both the EDM and the SM of hadrons and 
nuclei are difficult to calculate directly in QCD. 
However, long-range contributions
from pions can, to some extent, be calculated using the low-energy
effective field theory of QCD,
chiral perturbation theory (ChPT) \cite{weinberg79,Jenkins:1990jv,original}.
ChPT affords a systematic expansion of low-energy
observables in powers of $Q/M_{QCD}$, where $Q$ represents
low-energy scales such as external momenta and the pion mass $m_\pi$,
and $M_{QCD}\sim 1$ GeV denotes the characteristic QCD scale.
(For introductions, see for example Refs. \cite{Wbook,ulfreview}.)

In Refs. \cite{WHvK,jordyetal} the nucleon EDFF
stemming from $T$-violation sources of effective dimension up to 6
was considered in ChPT to the lowest order where momentum dependence
appears.
It was argued \cite{jordyetal} that the nucleon EDFF
partially reflects the source of $T$ violation at the quark level.
The various sources differ in particular in the expectation for
the behavior of the isoscalar EDFF.
For $\bar\theta$ and qCEDM, the isoscalar momentum dependence appears
only at NLO.
The nucleon EDFF from $\bar\theta$ was calculated at LO in 
Ref. \cite{WHvK}, generalizing to finite momenta earlier calculations 
of the EDM \cite{CDVW79,su3} and SM \cite{scott}. 
At this order, the momentum dependence is isovector
and completely due to a $T$-violating coupling of the pion 
cloud to the nucleon, with a radius fixed by $m_{\pi}^2$ \cite{scott}.
In Refs. \cite{willthesis,Ottnad} the EDFF calculation was extended to NLO,
and corrections found to be significant.
For the qCEDM, the nucleon EDFF was calculated at LO \cite{jordyetal},
at which order it is identical to that from $\bar\theta$.
For the other sources
of effective dimension 6, the quark EDM and the gluon color-EDM, 
the nucleon EDFF, including its isoscalar component,
was calculated to NNLO \cite{jordyetal} and found to be
mostly determined by short-distance physics.

Since the proposed deuteron experiment will probe the isoscalar
combination of neutron and proton EDMs (in addition to
$T$-violating two-nucleon effects),
we present here results for the nucleon EDFF to NLO from both
$\bar\theta$ and qCEDM,
using $SU(2)_L\times SU(2)_R$ heavy-baryon ChPT \cite{Jenkins:1990jv}.
For $\bar\theta$, we 
extend the calculation of Ref. \cite{willthesis}
and reproduce the EDM results
of Ref. \cite{Ottnad}, the latter obtained from a relativistic version of 
large-$N_c$ $U(3)_L\times U(3)_R$ ChPT, except for isospin-violating terms
neglected in Ref. \cite{Ottnad}. 
At this order, the isoscalar momentum dependence, and so the SM, is
entirely due to an isospin-breaking term related to the
nucleon mass splitting.
As we are going to see, no new undetermined parameters
appear, other NLO contributions being given by 
non-analytic recoil corrections proportional to $m_\pi/m_N$, 
where $m_N$ is the nucleon mass,
and by another isospin-breaking term, related to the pion mass splitting.
We use the non-analytic contributions to the isoscalar EDFF
to provide an estimate of the minimum expected size of
the deuteron EDM. 
The EDFF from the qCEDM depends at NLO  
on an additional $T$-violating pion-nucleon coupling,
although it is unlikely that the difference could be isolated experimentally.

For simplicity we focus here on QCD with two light quark flavors $u$ and $d$,
most relevant for low momenta $Q\sim m_{\pi}$, and consider as explicit degrees
of freedom only nucleons, pions, and photons.
In the framework of ChPT, the most general effective 
Lagrangian is 
built up using QCD symmetries as a guide,
in particular 
the chiral $SU_L(2)\times SU(2)_R\sim SO(4)$ symmetry,
which is spontaneously broken down to $SU(2)_{L+R}\sim SO(3)$.
A power-counting argument must be used to order interactions according to 
the expected size of their contributions.  
In order to fulfill chiral-symmetry requirements, pions couple derivatively 
in the chiral limit, which
brings to amplitudes powers of pion momenta.
Chiral-symmetry-breaking terms involve the quark masses $m_u$ and $m_d$,
so they bring into amplitudes powers of the pion mass.
Since nucleons are non-relativistic, we remove the large, inert nucleon
mass from nucleon fields \cite{Jenkins:1990jv}.
This gives one a chiral index ($\Delta$) with which to order terms 
in the Lagrangian \cite{weinberg79,original}, 
{\it i.e.} ${\cal L}=\sum_{\Delta}{\cal L}^{(\Delta)}$.  
For strong interactions,
the index is given by $\Delta=d+n/2-2$, where $n$ is the number of 
fermion fields and $d$ counts the number of derivatives and
powers of the pion mass.
Electromagnetic interactions are proportional to the small proton charge 
$e=\sqrt{4\pi\alpha_{em}}$, 
and it is convenient to account for factors of $e$ by enlarging the 
definition of $d$ accordingly.
Each interaction is associated with a parameter, or low-energy constant (LEC),
which can be estimated using naive dimensional analysis (NDA) 
\cite{NDA,Weinberg:1989dx}.
In this case, the index $\Delta$ tracks the number of inverse powers
of $M_{QCD}\sim 2\pi F_\pi\simeq 1.2$ GeV,
with $F_{\pi}\simeq 186$ MeV the pion decay constant,
associated with an interaction.
(Note that 
since NDA associates the LECs of chiral-invariant operators with $g_s/4\pi$,
for consistency one should take 
the strong-interaction coupling $g_s\sim 4\pi$.)

The theory can be enlarged in a straightforward way
to include the delta isobar.
Note that the delta isobar does not contribute to the nucleon
EDFF at the order in which we work.
As is the case for the nucleon anapole form factor \cite{AFF}, 
the structure of the delta interactions that would contribute at NLO vanishes 
in ChPT.  
The first nonvanishing delta contribution occurs at a higher order 
than we are considering here.

The $T$-conserving terms that we will need
consist of the following \cite{ulfreview,vanKolck2,vanKolck,willthesis}:
\begin{equation}
{\cal L}^{(0)}_{str/em}=
\frac{1}{2}D_{\mu}\boldpi \cdot D^{\mu}\boldpi
-\frac{m_{\pi}^{2}}{2D}\boldpi^{2}
+\bar{N}iv\cdot \mathcal D N
-\frac{2 g_{A}}{F_{\pi}} (D_{\mu}\boldpi) \cdot  \bar{N} \boldtau S^{\mu} N
\label{Lstr0}
\end{equation}
and
\begin{eqnarray}
{\cal L}^{(1)}_{str/em}&=& \frac{1}{2m_N}\left\{ 
-\bar{N}
\mathcal D_\perp^2 N
+\frac{2 g_A}{F_\pi} (i v \cdot D \boldpi ) \cdot \bar{N}  
 \boldtau S\cdot \mathcal D_{-} N                  
\right\}\nonumber \\
&&
+ \frac{e}{4m_N}\epsilon_{\mu\nu\rho\sigma} 
\bar{N}\left\{1+\kappa_0
+(1+\kappa_1) \left[\tau_3
  -\frac{2}{F_\pi^2 D}\left(\boldpi^2\tau_3-\pi_3\boldpi\cdot\boldtau\right)
  \right]\right\}v^\mu S^\nu N F^{\rho\sigma}
\nonumber \\
&& -\frac{\breve{\delta} m_\pi^2}{2D^2} \left(\boldpi^2 -\pi_3^2\right)
+ \frac{\delta m_N}{2} 
  \bar{N}\left(\tau_3-\frac{2 \pi_3}{F_\pi^2 D}\boldpi \cdot \boldtau \right)N.
\label{Lstr1}
\end{eqnarray}
Here $\boldpi$ denotes the pion field in a stereographic
projection of $SO(4)/SO(3)$, with $D=1+\boldpi^2/F_\pi^2$; 
$N=(p \; n)^T$ is a heavy-nucleon field of velocity $v^\mu$ and spin $S^\mu$
($S^\mu=(0, \vec{\sigma}/2$) in the nucleon rest frame where
$v^\mu=(1, \vec{0}$));
and $A_\mu$ is the photon field.
In addition, $(D_{\mu})_{ab} 
=D^{-1}(\delta_{ab}\partial_\mu -e\epsilon_{3ab}A_\mu)$
is the pion covariant derivative;
$\mathcal D_{\mu}=\partial_\mu 
+ i\boldtau\cdot(\boldpi\times D_{\mu}\boldpi)/F_\pi^2
-ieA_\mu(1+\tau_{3})/2$
is the nucleon covariant derivative;
and $F_{\mu\nu}=\partial_\mu A_\nu - \partial_\nu A_\mu$ 
is the electromagnetic field strength.
The component of $\mathcal D^{\mu}$ perpendicular to $v^\mu$
is written
\begin{equation}
\mathcal D_\perp^\mu =\mathcal D^\mu -v^\mu v\cdot \mathcal D,
\end{equation}
and we use the shorthand notation
\begin{equation}
\mathcal D_\pm^\mu \equiv \mathcal D^\mu \pm \mathcal D^{\dagger\mu} , 
\qquad 
\tau_i \mathcal D_\pm^\mu \equiv 
\tau_i\mathcal D^\mu \pm\mathcal D^{\dagger \mu} \tau_i,
\end{equation}
where ${\mathcal D}^\dagger$ is defined through
$\bar N \mathcal D^\dagger\equiv \overline{\mathcal D N}$.

The pion-nucleon coupling $g_A$
in Eq. (\ref{Lstr0}) and the anomalous magnetic photon-nucleon couplings
$\kappa_0$ and $\kappa_1$ in Eq. (\ref{Lstr1}) are not determined
by symmetry, but are expected to be ${\cal O}(1)$, and indeed
$g_{A}=1.3$, $\kappa_0=-0.12$, and $\kappa_1=3.7$. 
The pion mass term in Eq. (\ref{Lstr0}) originates
in explicit chiral-symmetry breaking by the average quark mass
$\bar{m}\equiv (m_d+m_u)/2$; from NDA,
$m_\pi^2={\cal O}(\bar{m} M_{QCD})$.
The contribution  to the nucleon mass from a similar term,
the nucleon sigma term, has been removed by an appropriate definition
of the heavy-nucleon field,
and the surviving nucleon-pion interactions do not contribute below.
The Goldberger-Treiman relation 
$g_{\pi NN}=2g_{A}m_{N}/F_{\pi}$  holds
in the two lowest orders, and a term in ${\cal L}^{(2)}_{str/em}$ provides an
${\cal O}(m_\pi^2/M_{QCD}^2)$ correction that accounts  \cite{vanKolck} for the
so-called Goldberger-Treiman discrepancy.

In Eq. (\ref{Lstr1}) we include explicitly
the leading isospin-breaking interactions \cite{vanKolck} stemming from
the quark mass difference, $m_d-m_u\equiv 2 \varepsilon \bar{m}$,
and from short-range electromagnetic effects.
The pion mass splitting is dominated by the electromagnetic
contribution $\breve{\delta} m_\pi^2= {\cal O}(\alpha_{em} M_{QCD}^2/4\pi)$;
because this is, numerically, of ${\cal O}(\varepsilon m_\pi^3/ M_{QCD})$ 
we book this term in ${\cal L}^{(1)}_{str/em}$.
The quark-mass contribution to the pion splitting is suppressed by a further 
$\varepsilon m_\pi/M_{QCD}$, one order down in the expansion.
Thus, to the accuracy in which we work here, $\breve{\delta} m_\pi^2$
is the observed pion mass splitting,
$\breve{\delta} m_\pi^2=1260$ MeV$^2$ \cite{pdg}.
Note that with the way we have written the splitting, in this
paper $m_\pi$ stands for the neutral pion mass,
the charged pion mass squared
being $m^2_{\pi^{\pm}} = m^2_\pi + \breve{\delta} m_\pi^2$.

The quark-mass contribution to the nucleon mass difference, $\delta m_N$,
is expected to be  ${\cal O}(\varepsilon m_\pi^2/M_{QCD})$
and it is evaluated to be 
$\delta m_N=2.26 \pm 0.57 \pm 0.42 \pm 0.10$ MeV 
from lattice simulations \cite{latticedeltamN},
which is in agreement with an extraction from charge-symmetry breaking 
in the $pn\to d \pi^0$ reaction \cite{CSBd}.
This is consistent with the NDA expectation that the corresponding 
electromagnetic contribution, $\breve{\delta} m_N$, is 
${\cal O}(\alpha_{em}M_{QCD}/4\pi)$ and thus somewhat smaller;
using the Cottingham sum rule \cite{Cott}, 
$\breve{\delta} m_N=-(0.76 \pm 0.30)$ MeV.
In our power counting, $\breve{\delta} m_N$ appears only in 
${\cal L}^{(2)}_{str/em}$.
It gives rise to different multi-pion interactions than does
the $\delta m_N$ term, but these multi-pion interactions do not matter below.
Thus, if desired, the dominant effect from $\breve{\delta} m_N$ can
be incorporated in our results with the simple substitution 
$\delta m_N \to \delta m_N + \breve{\delta} m_N$.
The isospin-breaking nucleon mass term can be eliminated with
simultaneous redefinitions of the pion and nucleon fields \cite{massplitt},
at the expense of new interactions. In the present case,
the relevant terms amount to adding to Eq. (\ref{Lstr1})
\begin{eqnarray}
\Delta{\cal L}^{(1)}_{str/em}&=& 
-\delta m_N\left(\boldpi \times v\cdot D\boldpi\right)_3
- \frac{\delta m_N}{2} 
  \bar{N}\tau_3 N.
\label{DeltaLstr1}
\end{eqnarray}
As a check, we have performed the calculation with and without 
Eq. (\ref{DeltaLstr1}), finding the same result, as required by 
field-redefinition invariance.

We consider two sources of $T$ violation at the scale $M_{QCD}$:
the QCD $\bar\theta$ term and the qCEDM. 
In terms of a gluon field strength $G^a_{\mu\nu}$
and an appropriate choice \cite{Baluni, BiraEmanuele} 
of quark fields $q=(u \; d)^T$, 
we can write them as
\begin{equation}
{\cal L}^{QCD}_{\slashT}=m_\star \bar{\theta} \; \bar{q}i\gamma_{5}q
- \frac{i}{2} \; \bar{q} \left(\tilde{d}_0 + \tilde{d}_3 \tau_3\right)
       \sigma^{\mu\nu}\gamma_{5}\lambda^a q \; G^a_{\mu\nu},
\label{TviolQCD}
\end{equation}
where 
\begin{equation}
m_\star= \frac{m_u m_d}{m_u+m_d}=\frac{\bar{m}}{2}(1-\varepsilon^2)
={\cal O}\left({\bar m}\right)
\end{equation}
and $\tilde{d}_0$ ($\tilde{d}_3$) is the isoscalar (isovector) qCEDM.
The first term in Eq. (\ref{TviolQCD})
represents the effect of the $\bar\theta$ term 
under the assumption that $\bar\theta$ is small, as inferred from the 
bound  \cite{dnbound} on the neutron EDM.
(For the more general case, see Ref. \cite{BiraEmanuele}.)
The second term in Eq. (\ref{TviolQCD})
is the QCD manifestation of  
sources of $T$ violation at a high scale $M_{\slashT}$.
At the Standard Model scale it is represented
by dimension-6 operators containing the mediator of 
electroweak-symmetry breaking \cite{Rujula,Weinberg:1989dx},
which at lower scales picks up a vacuum expectation value that
can be traded for $\bar{m}$.
We write \cite{jordyetal}
\begin{equation}
\tilde d_i=
{\cal O}\left(\frac{4\pi {\bar m}}{M_{\slashT}^2}{\tilde \delta}\right)
\end{equation}
in terms of a dimensionless factor
$\tilde{\delta}$.
The size of $\tilde{\delta}$ depends on the exact mechanisms of electroweak 
and $T$ breaking
and on the running to the low energies where non-perturbative QCD effects
take over.
The minimal assumption
is that it is ${\cal O}(g_s/4\pi)$, with 
$g_s$ the strong-coupling constant, 
but it can be 
much smaller (when parameters encoding $T$-violation beyond the Standard Model
are small) or much larger (since the first-generation
Yukawa couplings are unnaturally small).

The implications of $T$ violation to low-energy
observables depend on the way Eq. (\ref{TviolQCD}) breaks 
other QCD symmetries, in particular chiral symmetry 
\cite{willthesis,BiraEmanuele,morejordy}.
The $\bar\theta$ term 
is the fourth component of the same $SO(4)$ vector
$P=(\bar{q} \boldtau q, \bar{q} i\gamma_{5} q)$
that leads to isospin breaking \cite{WHvK, willthesis, BiraEmanuele}.
Therefore,
it generates EFT interactions that transform as $T$-violating,
fourth components
of $SO(4)$ vectors made out of hadronic fields,
with coefficients related to those of $T$-conserving interactions.
Similarly, the qCEDM breaks chiral symmetry as a combination
of fourth and third components of two other $SO(4)$ vectors 
\cite{willthesis,jordyetal, morejordy}.
As in the $T$-conserving case, we can use NDA to estimate 
the strength of the effective interactions,
and continue to label terms in
the effective chiral Lagrangian by the powers of $M_{QCD}^{-1}$.
Details of the construction of the Lagrangian from these terms are 
discussed in Refs. \cite{BiraEmanuele, morejordy}.

The relevant 
terms here
are the pion-nucleon interactions
\begin{equation}
{\cal L}^{(n)}_{\slashT}=
-\frac{1}{F_\pi D}\bar{N}
\left(\bar{g}_{0}\boldpi\cdot\boldtau 
+\bar{g}_{1} \pi_3 
\right) N
\label{LTviol-1}
\end{equation}
and
\begin{eqnarray}
{\cal L}^{(n+1)}_{\slashT} &=&
\frac{2}{F_\pi^2 D} (D_\mu\boldpi) \cdot \bar{N} \left(\bar{h}_{0}\boldpi 
+\bar{h}_{1}\pi_3 \boldtau \right) S^\mu N
\nonumber\\
&&
+\frac{\bar{h}_{2}}{F_\pi D} 
 \left(\delta_{i3}-\frac{2 \pi_i\pi_3}{F_\pi^2 D}\right) 
\bar{N} \left(\boldtau \times v\cdot D\boldpi\right)_i N,
\label{LTviol0}
\end{eqnarray}
and the short-range contributions to the nucleon EDM,
\begin{eqnarray}
{\cal L}^{(n+2)}_{\slashT} +{\cal L}^{(n+3)}_{\slashT} &=&
2\bar{N}\left\{\left(1-\frac{2 \boldpi^2}{F_\pi^2 D}\right)
                    \left[\bar{d}_{0}
                   +\bar{d}'_{1}
                    \left(\tau_{3}-\frac{2}{F_\pi^2 D}
        \left(\boldpi^2\tau_3-\pi_3 \boldpi\cdot\boldtau\right)\right)\right]
\right.
\nonumber\\
&& \qquad \left.
+\bar{d}_{1}\left(\tau_{3}-\frac{2 \pi_3}{F_\pi^2 D}
                    \boldpi\cdot\boldtau\right)\right\} 
S^{\mu}\left(v^{\nu}+\frac{i \mathcal D^{\nu}_{ \perp\, -}}{2 m_N}\right) N 
F_{\mu\nu}.
\label{LTviol23}
\end{eqnarray}
Here $\bar{g}_{i}$, $\bar{h}_{i}$, $\bar{d}_i$, and $\bar{d}'_{1}$
are parameters of sizes
\begin{equation}
\bar{g}_{0}= 
{\cal O}\left(\bar{\theta}\frac{m_\pi^2}{M_{QCD}},
              \tilde{\delta}\frac{m_\pi^2 M_{QCD}}{M_{\slashT}^2}\right),
\qquad 
\bar{g}_{1}= 
{\cal O}\left(\tilde{\delta}\frac{m_\pi^2 M_{QCD}}{M_{\slashT}^2}\right),
\label{NDA1}
\end{equation}
\begin{equation}
\bar{h}_{0}= 
{\cal O}\left(\bar{\theta}\frac{m_\pi^2}{M_{QCD}^2},
              \tilde{\delta}\frac{m_\pi^2}{M_{\slashT}^2}\right),
\qquad 
\bar{h}_{1,2}= 
{\cal O}\left(\tilde{\delta}\frac{m_\pi^2}{M_{\slashT}^2}\right),
\label{NDA2}
\end{equation}
and
\begin{equation}
\bar{d}_{0,1}, \bar{d}'_{1}= 
{\cal O}\left(e\bar{\theta}\frac{m_\pi^2}{M_{QCD}^3},
             e\tilde{\delta}\frac{m_\pi^2}{M_{\slashT}^2 M_{QCD}}\right).
\label{NDA3}
\end{equation}
The isoscalar ($\bar{d}_0$) and isovector ($\bar{d}_1$, $\bar{d}'_1$)
contributions to the nucleon EDM occur for both $T$-violation sources.
Direct  short-range contributions to the
momentum dependence of the EDFF first appear
in ${\cal L}^{(n+4)}_{\slashT}$,
being further suppressed by ${\cal O}(Q/M_{QCD})$. 
For $\bar\theta$, $n=1$ and among the $T$-violating $\pi N$ interactions
only the $I=0$ interactions with coefficients
$\bar{g}_{0}$ and $\bar{h}_0$ appear \cite{WHvK}.
Because of the link with isospin violation \cite{BiraEmanuele}, 
\begin{equation}
\bar{g}_0 = \frac{m_\star \delta m_N}{\varepsilon \bar{m}} \,\bar{\theta} 
\simeq  \frac{\delta m_N}{2\varepsilon} \, \bar{\theta}.
\label{g0constraint}
\end{equation}
A similar connection exists between 
$\bar{h}_0$ and the leading
isospin breaking in the pion-nucleon vertex.
For qCEDM, $n=-1$ and all $\pi N$ terms should be included.
Since in this case there is no analogous link to $T$-conserving quantities,
one cannot do better than the NDA estimates (\ref{NDA1}) and (\ref{NDA2})
without lattice or dynamical-model input.

The $T$-violating current-current nucleon-electron interaction is of the form
\begin{equation}
iT= -i e \, \bar{e}(l') \gamma^\mu e(l) \, D_{\mu\nu}(q)
        \, \bar{N}(p') J^\nu_{ed}(q,K) N(p),
\end{equation}
where $e(l)$ ($N(p)$) is an electron (nucleon) 
spinor with momentum $l$ ($p$)
and $D_{\mu\nu}(q)=-i(\eta_{\mu\nu}/q^2+\ldots)$ is the photon propagator 
with $q^2=(p-p^{\prime})^2\equiv -Q^2<0$.  
The nucleon electric dipole current $J_{ed}^{\mu}$ can be expressed 
in terms of $q=p-p^{\prime}$ and $K=(p+p')/2$
as an expansion in powers of $Q/m_N$ that 
reads \cite{WHvK,BiraEmanuele,jordyetal}
\begin{equation}
J^\mu_{ed}(q,K)= 2 
\left(F_0(Q^2)+F_1(Q^2) \tau_3\right)
\left[S^\mu v\cdot q- S\cdot q v^\mu
+\frac{1}{m_N}\left(S^{\mu} q\cdot K -S\cdot q K^{\mu}\right)+\ldots\right],
\label{cur}
\end{equation}
where $F_0(Q^2)$ ($F_1(Q^2)$)
is  the isoscalar (isovector) EDFF
of the nucleon. 
We will write
\begin{equation}
F_i(Q^2)=d_i - S'_i \; Q^2 + H_i(Q^2),
\label{Hdef}
\end{equation}
where 
$d_i$ is the EDM,
$S'_i$ the SM, and $H_i(Q^2)$ 
accounts for the 
remaining 
$Q^2$ dependence.

\begin{figure}[t]
\begin{center}
\scalebox{0.7}{\includegraphics{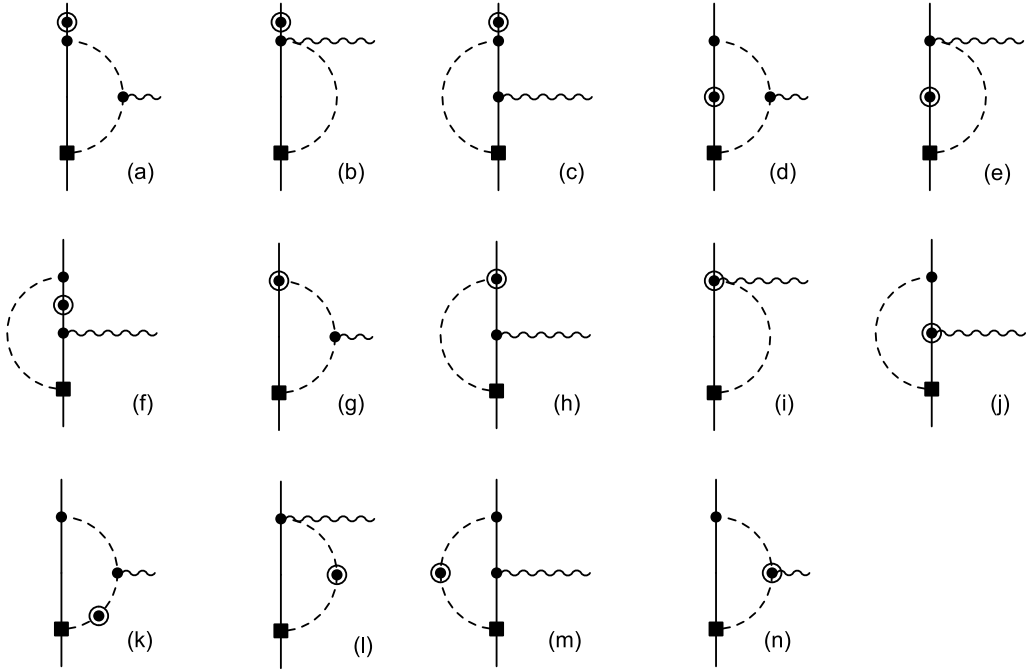}}
\end{center}
\caption{One-loop diagrams contributing to the nucleon
electric dipole form factor in sub-leading order coming from 
one insertion of an ${\cal L}^{(1)}_{str/em}$ operator.  
Solid, dashed and wavy lines represent nucleons, pions and (virtual) photons, 
respectively; 
single filled circles stand for interactions from ${\cal L}^{(0)}_{str/em}$
while double circles for interactions from ${\cal L}^{(1)}_{str/em}$ 
and/or 
$\Delta \mathcal L^{(1)}_{str/em}$;
squares represent the $T$-violating vertices from ${\cal L}^{(n)}_{\slashT}$.
For simplicity only one possible ordering is shown here.}
\label{edfig1}
\end{figure}

\begin{figure}[t]
\begin{center}
\scalebox{0.7}{\includegraphics{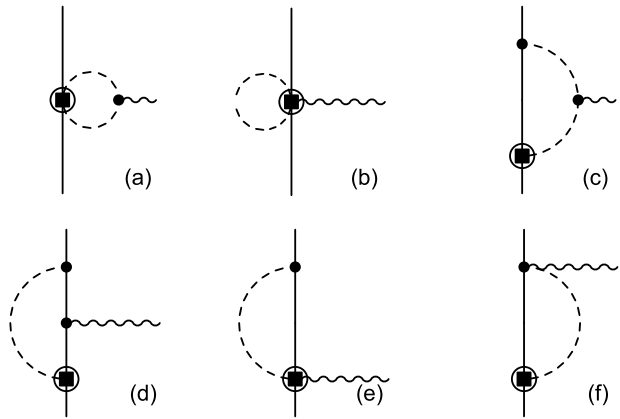}}
\end{center}
\caption{Diagrams contributing to the nucleon electric dipole form factor 
in sub-leading order coming from 
one insertion of the $T$-violating vertex from ${\cal L}^{(n+1)}_{\slashT}$, 
represented by a circled square. Other symbols are as in Fig. \ref{edfig1}.
For simplicity only one possible ordering is shown here.}
\label{edfig2}
\end{figure}

The form factor itself can be expanded in powers
of $Q/M_{QCD}$. The leading-order (LO) contributions to the current,
which are ${\cal O}(e\bar{g}_0 Q/(2\pi F_{\pi})^2)$,
have been 
calculated in Refs. \cite{WHvK,jordyetal}.
They include the unknown short-range contributions
in ${\cal L}^{(n+2)}_{\slashT}$ (\ref{LTviol23}), and loop diagrams
made out of $T$-violating interactions in 
${\cal L}^{(n)}_{\slashT}$ (\ref{LTviol-1})
and $T$-conserving interactions in ${\cal L}^{(0)}_{str/em}$ (\ref{Lstr0}).
Here we focus on 
the next-to-leading order (NLO),
that is, terms of relative ${\cal O}(Q/M_{QCD})$.
They are made of diagrams with one insertion of
interactions in 
${\cal L}^{(n+3)}_{\slashT}$ (\ref{LTviol23}),
${\cal L}^{(n+1)}_{\slashT}$ (\ref{LTviol0}),
or
${\cal L}^{(1)}_{str/em}$ (\ref{Lstr1}).
There are no new, unknown short-range parameters appearing at tree level,
the recoil corrections
in ${\cal L}^{(n+3)}_{\slashT}$ (\ref{LTviol23})
simply ensuring 
---together with those in ${\cal L}^{(1)}_{str/em}$ (\ref{Lstr1})--- 
the form (\ref{cur}) of the current.
The loop diagrams contributing to the nucleon EDFF in NLO are
shown in Figs. \ref{edfig1} and \ref{edfig2},
classified according to the combination of couplings that they contain.
All other contributions to the EDFF are formally of
higher order: they come from more powers of momenta in diagrams with
the same number of loops, or from extra loops.

The NLO diagrams of Fig. \ref{edfig1} are built from the leading interactions 
in Eqs. (\ref{Lstr0}) and (\ref{LTviol-1}), 
plus one insertion of an operator from Eq. (\ref{Lstr1}).  
Diagrams \ref{edfig1}(a,b,c) represent a correction 
to the external energies,
\begin{eqnarray}
v\cdot q &=& -\frac{q\cdot K}{m_N},
\label{onshell1}
\\
v\cdot K &=& -\frac{1}{2m_N}\left(K^2 +\frac{q^2}{4}\right)
             \mp \frac{\delta m_N}{2},
\label{onshell2}
\end{eqnarray}
of a proton ($-$ sign) or neutron ($+$ sign) in LO diagrams. 
(In the remaining NLO diagrams, we set the right-hand side of these 
equations to zero.)
Analogous insertions 
in the nucleon propagator are represented by diagrams \ref{edfig1}(d,e,f).
Diagrams \ref{edfig1}(g,h,i) originate in the recoil correction in pion 
emission/absorption, while diagram \ref{edfig1}(j) arises
from the magnetic photon-nucleon interaction.  
Diagrams \ref{edfig1}(k,l,m) represent an insertion
of the pion (and
nucleon, in the case of Eq. \eqref{DeltaLstr1})
mass splitting in pion propagation, while diagram \ref{edfig1}(n) 
contains the isospin-violating photon-pion interaction induced by the 
first term of Eq. \eqref{DeltaLstr1}.
These one-loop diagrams contribute to the current at order 
${\cal O}(e\bar{g}_i Q^2/(2\pi F_{\pi})^2 m_N)$.  

The NLO diagrams in Fig. \ref{edfig2} are built from the leading interactions 
in Eq. (\ref{Lstr0}) with one insertion of an operator from 
Eq. (\ref{LTviol0}).  
Diagrams 2(a,b) stem from the sub-leading pion-nucleon 
couplings $\bar{h}_{0,1}$, and 
diagrams 2(c,d,e,f) from the sub-leading coupling $\bar{h}_{2}$,
present only for qCEDM.
These one-loop diagrams contribute to the current at order 
${\cal O}(e\bar{h}_i Q^2/(2\pi F_\pi)^2)$, 
which is precisely the same order as the diagrams in Fig. \ref{edfig1}.  

The diagrams in Figs. \ref{edfig1} and \ref{edfig2} can be evaluated
in a straightforward way.
We use regularization in $d$ spacetime dimensions, 
and define
\begin{equation}
L \equiv \frac{2}{4-d} -\gamma_{E}+\ln 4\pi,
\end{equation}
where $\gamma_E=0.557\ldots$ is the Euler constant. 
The LO loop contributions depend on a renormalization scale $\mu$
but this dependence is compensated for by the contribution
from the short-range interactions in Eq. (\ref{LTviol23}).
The NLO diagrams are finite in this regularization scheme.

Most of the diagrams actually vanish when the on-shell conditions
(\ref{onshell1}) and (\ref{onshell2}) are consistently enforced.
Diagrams (a,b) in Fig. \ref{edfig2}
vanish due to isospin. 
Since diagrams \ref{edfig2}(c,d,e,f)
vanish too, 
the EDFF to this order depends only on the leading $T$-violating parameters
$\bar g_{0,1}$ through Fig. \ref{edfig1}.
Diagram \ref{edfig1}(j) vanishes due to its spin structure and
therefore the EDFF does not depend on the anomalous magnetic moments, either.
Diagram \ref{edfig1}(h)  gives both isoscalar and isovector contributions. 
The remaining non-vanishing diagrams are \ref{edfig1}(a,d,k).
Neglecting $T$-conserving isospin violation,
these diagrams give purely isovector results.
In the case of $\bar \theta$, the results 
are proportional to $eg_A\bar{g}_0/(2\pi F_\pi)^2$,
as in LO \cite{WHvK}, times the recoil suppression factor $m_\pi/m_N$.
For qCEDM, there is an additional momentum-independent
contribution proportional to $\bar g_1$.

We have checked each of the
isospin-breaking contributions in two ways.
The contributions from ${\breve \delta}m_\pi^2$ come through 
diagrams \ref{edfig1}(k,l,m). 
Because the LO EDFF originates entirely in charged-pion diagrams,
these contributions can be obtained
alternatively by evaluating the LO EDFF with $m_\pi^2+{\breve \delta}m_\pi^2$,
then expanding in powers of ${\breve \delta}m_\pi^2/m_\pi^2$:
\begin{equation}
F_1(Q^2)|_{m_\pi^2+{\breve \delta}m_\pi^2}=
F_1(Q^2)|_{m_\pi^2}
+{\breve \delta}m_\pi^2 
\frac{\partial F_1(Q^2)}{\partial m_\pi^2}|_{m_\pi^2} +\ldots
\end{equation} 
The resulting EDFF is thus isovector.
Including the nucleon mass difference $\delta m_N$, diagrams 
\ref{edfig1}(a,d)
generate an additional isoscalar contribution. 
If we instead use the field redefinition of Ref. \cite{massplitt},
we drop the $\delta m_N$ term in Eq. (\ref{onshell2})
and in internal nucleon lines.
The same result is obtained from
extra contributions
$\propto \delta m_N$ in diagrams \ref{edfig1}(k,l,m),
and a new 
isospin-breaking photon-pion coupling,
depicted in diagram \ref{edfig1}(n). 

The diagrams in Fig. \ref{edfig1} contribute 
to both isoscalar and isovector EDMs. 
Taking the NLO contributions together with the LO 
from Refs. \cite{WHvK,jordyetal}, we have
\begin{eqnarray}
d_0 &=& \bar{d}_0
        +\frac{eg_A\bar{g}_0}{(2\pi F_{\pi})^2}\; \pi
        \left[\frac{3m_{\pi}}{4m_N}
              \left(1+\frac{\bar{g}_1}{3\bar{g}_0}\right)
              -\frac{\delta m_N}{m_{\pi}}
        \right],
\label{d0}\\
d_1 &=& \bar{d}_1 + \bar{d}'_1
        +\frac{eg_{A}\bar{g}_{0}}{(2\pi F_{\pi})^2}
        \left[L-\ln\frac{m_{\pi}^2}{\mu^2}
        +\frac{5\pi}{4}\frac{m_\pi}{m_N}
         \left(1+\frac{\bar{g}_1}{5\bar{g}_0}\right)
        -\frac{\breve{\delta} m_\pi^2}{m_{\pi}^2}\right].
\label{d1}
\end{eqnarray}
The LO piece in Eq. (\ref{d1}), which depends on $\bar{g}_0$
and is non-analytic in $m_\pi^2$,
is, with the use of the Goldberger-Treiman relation,
the result of Ref. \cite{CDVW79},
which holds also for the qCEDM \cite{jordyetal}.
The short-range isovector 
combination $\bar{d}_1 + \bar{d}'_1$
absorbs the divergence and $\mu$ dependence of the LO loop.
The short- and long-range contributions to the EDM
are in general of the same size, but a cancellation
is unlikely due to the non-analytic dependence on $m_\pi$
of the pion contribution.
The isoscalar parameter $\bar{d}_0$ is not needed for renormalization
at this order,
but there is no 
apparent reason to assume its size to be 
much smaller than NDA either.

At NLO, the EDM receives finite non-analytic corrections,
which depend also on $\bar{g}_1$ for qCEDM.
{}From Eqs. (\ref{d0}) and (\ref{d1}) we see that, as usual in baryon
ChPT, the NLO contributions are enhanced by $\pi$ over NDA.
However, the other dimensionless factors are not large enough
to overcome the $m_\pi/m_N$ suppression.
Setting $\mu$ to $m_{N}$ as a representative value
for the size of $d_1$ \cite{CDVW79}, 
the NLO term in Eq. (\ref{d1}) (Eq. (\ref{d0})) is about 15\% 
(10\%) 
of the leading  non-analytic term in Eq. (\ref{d1}),
indicating good convergence of the chiral expansion. 
The isovector character of the LO non-analytic terms is approximately
preserved at NLO.
Isospin-breaking contributions, although formally NLO,
are pretty small, amounting to 15-20\% of the total NLO contribution.

In the case of $\bar\theta$ 
we can use Eq. (\ref{g0constraint}) and expect
\begin{eqnarray}
|d_n| = |d_0 -d_1| &\simge& 
\frac{eg_{A}}{(2\pi F_{\pi})^2}\frac{\delta m_N}{2 \varepsilon} 
             \left[ \ln\frac{m_N^2}{m^2_{\pi}} 
             + \frac{\pi}{2} \frac{m_{\pi}}{m_N} 
             - \frac{\breve{\delta}m^2_{\pi}}{m^2_{\pi}} 
             + \pi \frac{\delta m_N}{m_{\pi}}  \right] \bar\theta 
\nonumber\\
&\simeq& (1.99 + 0.12 - 0.04 + 0.03 ) \cdot 10^{-3} 
         \; \bar\theta \; e \, {\rm fm} 
\end{eqnarray}
for the neutron EDM 
and
\begin{eqnarray}
|d_p| = |d_0+d_1|&\simge& 
\frac{eg_{A}}{(2\pi F_{\pi})^2}\frac{\delta m_N}{2 \varepsilon} 
            \left[ \ln\frac{m_N^2}{m^2_{\pi}} 
             + 2\pi \frac{m_{\pi}}{m_N} 
             - \frac{\breve{\delta}m^2_{\pi}}{m^2_{\pi}} 
             - \pi \frac{\delta m_N}{m_{\pi}}  \right] \bar\theta 
\nonumber\\
&\simeq& (1.99 + 0.46 - 0.04 - 0.03 ) \cdot 10^{-3} 
         \; \bar\theta \; e \, {\rm fm} 
\end{eqnarray}
for the proton EDM, 
using the lattice QCD value 
$\delta m_N/2\varepsilon = 2.8$ MeV \cite{latticedeltamN}.
Non-analytic NLO corrections are therefore somewhat larger for the proton EDM,
but this difference is unlikely to be significant in light of our ignorance
about the size of short-range contributions.

The non-analytic terms in Eq. (\ref{d0}) 
represent a lower-bound estimate for the size of the nucleon isoscalar EDM,
as the short-range contribution $\bar{d}_0$ is nominally of lower order.
The expected lower bound on the  nucleon isoscalar EDM might have implications
for proposed experiments on EDMs of light nuclei.
In these cases, there will be additional many-nucleon contributions,
but the average of the one-nucleon contributions still provides an estimate 
of the order of magnitude of the expected nuclear EDM.
For the deuteron, the average 
one-nucleon contribution is exactly $d_0$ and, in the case of $\bar \theta$, 
we expect for the 
deuteron EDM 
\begin{equation}
\left|d_d\right| \simge \frac{e g_A}{\left(2\pi F_{\pi}\right)^2} 
                 \frac{\delta m_N}{2\varepsilon} \, \pi 
                 \left[\frac{3m_{\pi}}{4m_N} 
                 - \frac{\delta m_N}{m_{\pi}}\right] \bar\theta 
\simeq ( 1.7 - 0.3 )\cdot 10^{-4} \, \bar\theta \, e\, \textrm{fm} .
\label{deuteron}
\end{equation}
Therefore, if 
there are no cancellations, a deuteron EDM signal from $\bar\theta$
is expected to 
be larger than about 10\% of the neutron EDM signal. 

Note that short- and long-range physics cannot
be separated with a measurement of the neutron and proton EDMs alone. 
On the other hand, the momentum dependence of the EDFF
is completely determined, to the order we are working, 
by long-range contributions generated by $\bar{g}_0$.
It is therefore the same for $\bar\theta$ and qCEDM.
It turns out that the isoscalar form factor receives momentum dependence
only from isospin-breaking terms, 
while there is a non-vanishing correction to the
isovector momentum dependence also from isospin-conserving terms.

The variation of the form factor with $Q^2$ can be characterized
at very small momenta by the electromagnetic contribution to the nucleon SM, 
the leading  and sub-leading contributions of which we find to be 
\begin{eqnarray}
S'_0&=& -\frac{eg_A\bar{g}_0}{6(2\pi F_{\pi})^2 m_{\pi}^2}
       \frac{\pi}{2}\frac{\delta m_N}{m_{\pi}},
\\
S'_1&=&\frac{eg_A\bar{g}_0}{6 (2\pi F_{\pi})^2 m_{\pi}^2}
       \left[1-\frac{5\pi}{4}\frac{m_{\pi}}{m_N}
             -\frac{\breve{\delta} m_\pi^2}{m_{\pi}^2} \right].
\label{radius}
\end{eqnarray}
The LO, isovector term is the result of Refs. \cite{scott,WHvK}.
While the EDM vanishes in the chiral limit, the isovector SM is finite.
The NLO correction, which agrees with the $\bar\theta$ result
of Ref. \cite{Ottnad} when $T$-conserving isospin violation is neglected,
vanishes in the chiral limit but 
gives a relatively large correction to the isovector SM of about 60\%,
due to the numerical factor $5\pi/4$.
Again, the isospin-breaking corrections are relatively small,
and, as a consequence, at NLO the SM remains
mostly isovector.

To this order, the SM is entirely given, apart for $\bar{g}_0$, by
quantities that can be determined from other processes. 
In the case of $\bar\theta$, we can again use 
Eq. (\ref{g0constraint}) to estimate
\begin{eqnarray}
S'_0&=&
-\frac{eg_A}{12(2\pi F_{\pi})^2}
\frac{\pi (\delta m_N)^2}{2\varepsilon m_{\pi}^3}\, \bar\theta
\simeq  - 5.0 \cdot 10^{-6} \, \bar\theta\; e\,{ \rm fm}^3,
\label{SM0}
\\
S'_1&=&
\frac{eg_A}{12 (2\pi F_{\pi})^2}\frac{\delta m_N}{\varepsilon m_{\pi}^2}
\left[1-\frac{5\pi}{4}\frac{m_{\pi}}{m_N}
-\frac{\breve{\delta} m_\pi^2}{m_{\pi}^2}\right] \bar\theta 
\simeq 6.8 \cdot 10^{-5} \, \bar\theta\; e\, {\rm fm}^3,
\label{radiusestimate}
\end{eqnarray}
where again we used the lattice-QCD value \cite{latticedeltamN} for
$\delta m_N/2\varepsilon$.
{}From these results we can straightforwardly obtain the SM for the proton and
the neutron.
Although we could again use the isoscalar component 
as an estimate for a lower bound on the deuteron SM,
there could be potentially significant contributions from the deuteron
binding momentum. 

The full momentum dependence of the EDFF is given in addition
by the functions $H_i(Q^2)$ introduced in Eq. (\ref{Hdef}),
\begin{eqnarray}
H_{0}(Q^{2})&=&-\frac{4eg_A\bar{g}_0}{15(2\pi F_{\pi})^2}\, \frac{3\pi}{4}
                \frac{\delta m_N}{m_{\pi}}
                \; h_0^{(1)}\left(\frac{Q^2}{4m_\pi^2}\right),
\\
H_{1}(Q^{2})&=&\frac{4eg_{A}\bar{g}_{0}}{15(2\pi F_{\pi})^{2}}
               \left[h_1^{(0)}\left(\frac{Q^{2}}{4m_{\pi}^{2}}\right)
                     -\frac{7\pi}{8}\frac{m_{\pi}}{m_N}
                      \; h_1^{(1)}\left(\frac{Q^2}{4m_\pi^2}\right)
                     -\frac{2\breve{\delta} m_\pi^2}{ m_{\pi}^2}
                  \;{\breve h}_1^{(1)}\left(\frac{Q^2}{4m_\pi^2}\right)\right].
\label{FD}
\end{eqnarray}
Here, the LO 
term,
\begin{equation}
h_1^{(0)}(x)=-\frac{15}{4}\left[
        \sqrt{1+\frac{1}{x}} \;
        \ln{\left(\frac{\sqrt{1+1/x}+1}{\sqrt{1+1/x}-1}\right)}
        -2\left(1+\frac{x}{3}\right)\right],
\label{f0}
\end{equation}
is the one calculated in Refs. \cite{WHvK,jordyetal},
while we now obtain the NLO isovector functions 
\begin{equation}
h_1^{(1)}(x)
=-\frac{1}{7}\left[3(1+2x)\, h_0^{(1)}(x)-10x^2\right],
\label{f1}
\end{equation}
and 
\begin{equation}
{\breve h}_1^{(1)}(x)
=-\frac{1}{4(1+x)}
  \left(h_1^{(0)}(x)-5x^2\right),
\label{f3}
\end{equation}
and the NLO isoscalar function
\begin{equation}
h_0^{(1)}(x)=5\left(
\frac{1}{\sqrt{x}} \arctan \sqrt{x} -1 +\frac{x}{3}\right).
\label{f2}
\end{equation}
In compliance with the definition of $H_i$, the four functions 
behave as
$h_i^{(n)}(x)=x^2 +{\cal O}(x^3)$ for $x\ll 1$.

\begin{figure}[t]
\begin{center}
\scalebox{0.5}{
\includegraphics{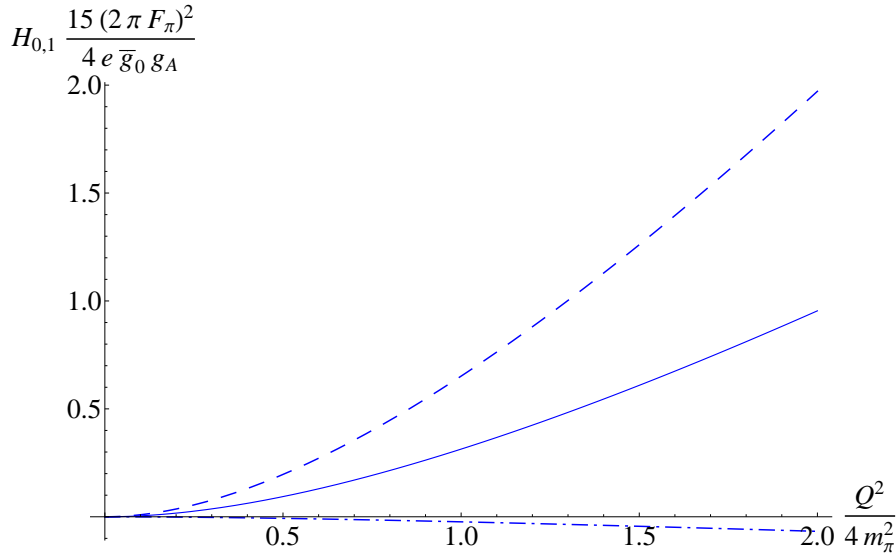} 
}
\end{center}
\vspace{-0.5cm}
\caption{The isovector $H_{1}(Q^2)$  in LO
(dashed line) and LO+NLO (solid line),
and the isoscalar $H_0(Q^2)$ in LO+NLO (dash-dotted line),
both in units of $4e g_A \bar g_0/15 \, (2\pi F_{\pi})^2$,
as functions of $Q^2$ , in units of $4m_\pi^2$.  
}
\label{edfig3}
\end{figure}

As in lowest order, the momentum dependence is fixed by the pion cloud.  
Thus the scale for momentum variation is determined by $2m_\pi$.
Both the SM and the functions $H_{0,1}(Q^{2})$
are testable predictions 
of ChPT.
In Fig. \ref{edfig3}
we plot 
the LO $h_1^{(0)}$, 
the LO+NLO combination 
$h_1^{(0)}-(7\pi m_\pi/8m_N)h_1^{(1)}
-(2\breve{\delta} m_\pi^2/ m_{\pi}^2){\breve h}_1^{(1)}$,
and the NLO $-(3\pi \delta m_N/4m_{\pi})h_0^{(1)}$ as 
functions of $Q^2$.
We use the same values of parameters as before.
As for the SM, NLO corrections can be significant, but the 
isospin-breaking contributions are small.

In summary, 
we have calculated the 
nucleon electric dipole form factor due to the $\bar\theta$ term 
and to the quark color-electric dipole moment
in sub-leading order in ChPT, including isospin-breaking effects.
The chiral expansion seems to be converging,
although NLO corrections are enhanced by extra factors of $\pi$.
Under the assumption that higher-order results are not afflicted by
anomalously-large dimensionless factors,
the relative error of our results
at momentum $Q$ should be $\sim (Q/M_{QCD})^2$.
The NLO isospin-breaking contributions are relatively small 
and could be overcome by isospin-conserving contributions at NNLO.
We have shown that at NLO the EDFF includes both isoscalar and isovector
components, with a $Q^2$ dependence determined by
non-derivative $T$-violating pion-nucleon couplings
and the pion mass.
The isoscalar momentum dependence is entirely due
to the nucleon mass splitting.
We have provided 
a lower-bound estimate for the isoscalar nucleon EDM,
expected to set also the minimum size of the deuteron EDM.
A full calculation of the latter in ChPT can now be performed.

\vspace{0.5cm}
\noindent
{\bf Acknowledgments}

\noindent
We thank R. Timmermans for stimulating discussions.
We acknowledge the hospitality 
of the Department of Physics at the University of Arizona (CMM),
the Theory Division at CERN (EM),
the Kernfysisch Versneller Instituut at the Rijksuniversiteit Groningen 
(EM, UvK),
and the Nuclear Theory Group at the University of Washington (UvK),
while this work was being carried out.  
This research was supported in part by 
an APS Forum of International Physics travel grant (CMM, UvK),
by FAPERGS-Brazil under contract PROADE 2 02/1266.6 (CMM),
by the Brazilian CNPq under contract 474123/2009 (CMM),
by the Dutch Stichting voor Fundamenteel
Onderzoek der Materie under programme 104 (JdV),
by the US Department of Energy under
grants DE-FG02-04ER41338 (EM, WHH, UvK)	
and DE-FG02-06ER41449 (EM),
and by the Alfred P. Sloan Foundation (UvK).

\end{document}